\documentclass[a4paper, oneside, twocolumn, notitlepage, 10pt]{extarticle_ecoc}
\usepackage{ecoc}

\newcommand{\abs}[1]{\lvert#1\rvert}

\begin{document}
\selectlanguage{english}    


\title{Field-Enhanced Filtering in MIMO Learned Volterra Nonlinear Equalisation of Multi-Wavelength Systems}%

\author{
    Nelson Castro\textsuperscript{(1,$\ast$)}, Sonia Boscolo\textsuperscript{(1)},
    Andrew D. Ellis\textsuperscript{(1)}, Stylianos Sygletos\textsuperscript{(1)}
}

\maketitle                  


\begin{strip}
    \begin{author_descr}

        \textsuperscript{(1)}Aston Institute of Photonic Technologies, Aston University, Birmingham, UK,
        \textsuperscript{($\ast$)}\textcolor{blue}{\uline{cast1901@aston.ac.uk}} 

    \end{author_descr}
\end{strip}

\renewcommand\footnotemark{}
\renewcommand\footnoterule{}

\begin{strip}
    \begin{ecoc_abstract}
        We propose a novel MIMO-WDM Volterra-based nonlinear-equalisation scheme with adaptive time-domain nonlinear stages enhanced by filtering in both the power and optical signal waveforms. This approach efficiently captures the interplay between dispersion and non-linearity in each step, leading to $46\%$ complexity reduction for $9\times 9$-MIMO operation.
        
        \textcopyright2024 The Author(s)
    \end{ecoc_abstract}
\end{strip}


\section{Introduction}
Digital compensation of nonlinear transmission impairments offers a promising path to increase the capacity of optical communication systems \cite{Ellis:APC2022}. 
However, effective compensation of inter-channel impairments, which dominate in wavelength division multiplexing (WDM) transmission, requires multi-channel operations within the equaliser, significantly increasing its computational complexity.
Over the last few years, a number of research efforts have aimed to reduce the associated cost and make multi-channel nonlinear equalisation (NLE) commercially viable \cite{Mateo:OE2010, maher_spectrally_2015}. In particular, the development of multiple-input multiple-output (MIMO) digital back-propagation (DBP) based models \cite{Mateo:OE2010, Civelli:ISWCS21}, utilising enhanced split-step Fourier (SSF) methods, has significantly reduced the number of required computational steps per span and  the algorithmic complexity compared to full-field wide-band DBP-based approaches \cite{Liga:OE2014}. 
The enhanced approach consisted in the application of linear MIMO 
filtering  to the power waveform of co-propagating signal channels at each nonlinear stage of the SSF algorithm in the frequency or time domain \cite{Mateo:OE2010, Secondini:PNC2016, Civelli:ISWCS21}. 
The recent integration of machine learning (ML) into NLE has made a major impact. Utilising gradient back-propagation tools has enabled more effective optimisation of the algorithm's parameters, substantially improving the equalisation capability and further reducing the algorithmic complexity\cite{Hager:JSAC2021}.  ML has significantly improved the efficiency of both DBP and Inverse Volterra Series Transfer Function (IVSTF)-based MIMO NLE schemes, enabling their operation at a single computational step per span for long-haul transmission \cite{Sidelnikov:JLT2021,Castro:ONDM2024, Castro:ACP23}, through the joint optimisation of the algorithm's parameters in the linear and nonlinear stages. However, these fully time-domain (TD) models require finite-impulse-response (FIR) filters to compensate the chromatic dispersion (CD) at each linear stage with lengths that are not commercially available yet, i.e., $\sim 40$ taps for $\sim 50$-km computational step length at $32$-Gbaud rate.
\begin{figure}[t!]
    \centering
    \includegraphics[height=6.5cm]{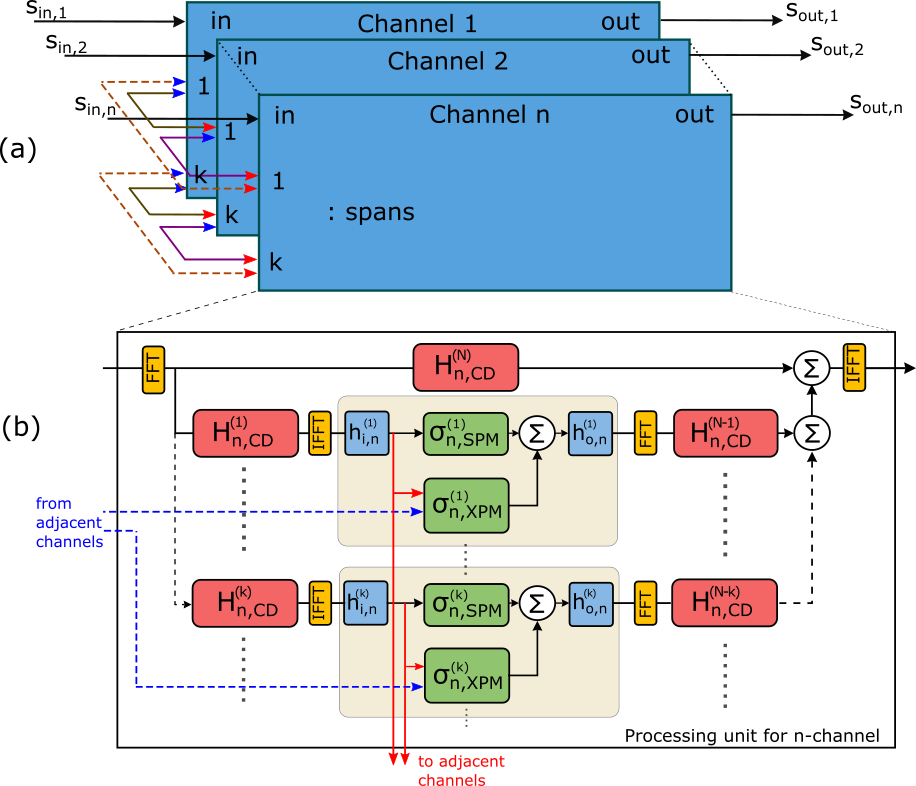}
    \caption{Block diagram of the proposed MIMO learned IVSTF NLE. (a) Interconnection of different channel's units. (b) Processing unit corresponding to the $n^{\text{th}}$ channel. Bi-directional arrows between adjacent channel steps show fields directed to and from XPM activation functions.}
    \label{fig:figure1}
\end{figure}

Handling CD in the frequency domain (FD) using static equalisers embedded in ASIC units is computationally simpler and offers a more energy-efficient implementation for long-haul transmission scenarios. However, opting for static operation in the linear stages of the NLE, while maintaining adaptive nonlinear stages only, may compromise the equalisation effectiveness. While initial studies have 
been conducted for single- and multi- channel equalisation scenarios \cite{Secondini:PNC2016, Civelli:ISWCS21}, they could not maximise the performance or complexity gain as they used brute-force optimisation within a limited parameter space that was the same across all nonlinear stages. 
In our recent work \cite{Castro:CLEO2024}, the joint optimisation of the FIR parameters of filtered nonlinear stages afforded by ML optimisation methods enabled effective operation of IVSTF-based multi-channel equalisers for MIMO sizes of up to $7\times 7$ and $9\times 9$. Yet, the model required four steps per span to equal the performance of its fully TD counterpart \cite{Castro:ONDM2024}. It is desirable to reduce further the required number of computational steps.

In this paper, we propose a field-enhanced (FE) filtering scheme for the nonlinear stages of an IVSTF-based model,
which restores the adaptive operation advantage of its purely TD counterpart, without compromising its practical realisation potential. By incorporating short ($3$-tap) FIR filters into both the input and output of each nonlinear stage 
to filter the signal's field waveform, we demonstrate a significant improvement of the equalisation capability of the algorithm.
Our results demonstrate successful operation of the FE L-IVSTF NLE scheme at only two steps per span for a $9\times 9$ MIMO configuration, matching the performance of L-IVSTF systems with a twice larger step count while reducing the computational complexity by $46\%$. 


\section{Proposed Equalisation Scheme}
The equaliser's structure, based on the parallel IVSTF scheme proposed in \cite{liu_intrachannel_2012},  is shown in Fig.\ \ref{fig:figure1}. 
An FD filter addresses the CD accumulated over the entire transmission link, while the nonlinear distortions are simultaneously estimated in the parallel branches. In the $k^{\text{th}}$ branch, the first linear stage manages the CD up to the point $z = kL_{\mathrm{sp}}/N_{\mathrm{StpS}}$ ($L_{\mathrm{sp}}$ is the fibre span's effective length, $N_{\mathrm{StpS}}$ is the number of steps per span), localising there the $k^{\text{th}}$ step's nonlinear phase shift. A subsequent linear stage compensates the CD up to the link's end. Each FD filter used in the linear stages includes an appropriate term to address the walk-off effect. Between the linear stages, nonlinear transformations account for the effects of self-phase modulation (SPM) and cross-phase modulation (XPM) to mitigate the nonlinearity within each step. Our proposed FE filtering approach outfits the input/output ports of each nonlinear branch with a pair of short CD FIR filters, identical in length. These filters operate on the complex field waveform of each optical channel $n$, addressing a predefined amount of dispersion not covered by the FD filters. Notably, these filters complement the FIR filters within enhanced MIMO nonlinear stages\cite{Castro:CLEO2024},
which manage the signal power waveform as a part of the SPM: $ {\sigma}_{n,\mathrm{SPM}}^{(k)}(t) = -j\gamma L_{\mathrm{sp}}  y_n^{(k)}(t)\sum_{c=-l}^l \alpha_{c}^{(k)} \abs{y_n^{(k)}(t+cT_s)}^{2}$, and of the XPM: $\sigma_{n,\mathrm{XPM}}^{(k)}(t)=  -2j \gamma L_{\mathrm{sp}} y_n^{(k)}(t)\sum_{r\not=n}^{N_{\mathrm{ch}}}\sum_{c=-m}^m \beta_{c,r}^{(k)} \abs{y_r^{(k)}(t+cT_s)}^2$ activation functions, where $y_n^{(k)}(t)$ is the complex field of the channel, $\gamma$ is the  nonlinear coefficient of the fibre, $\alpha_c^{(k)}$ and $\beta_{c,r}^{(k)}$ are the FIR filter’s coefficients, $T_s$ is the sampling interval, and $N_{\mathrm{ch}}$ is the dimension of the MIMO algorithm. The scheme features an optimised placement of the required fast Fourier transforms (FFTs) \cite{liu_intrachannel_2012}. 


\section{Simulation Setup, Results and Discussion}
In this study, we simulated the transmission of $11$ single-polarisation wavelength channels over a $6 \times 100$-km standard single-mode fibre link (dispersion parameter $D = 17\,\mathrm{ps/(nm \cdot km)}$, $\gamma = 1.3\,\mathrm{(W \cdot km)}^{-1}$, loss coefficient $\alpha = 0.2\,\mathrm{dB/km}$). Erbium-doped fibre amplifiers of $4.5\,\mathrm{dB}$ noise figure compensated for the span losses. Each channel was modulated with $64$ quadrature-amplitude modulation symbols at a rate of $32\,$Gbaud. The channel spacing was $40\,$GHz.  At the receiver, the channels of interest were demultiplexed and down-sampled to $2$ samples per symbol before being processed by the MIMO NLE. Following the NLE stage, each channel was matched filtered and further down-sampled to $1$ sample per symbol. The DSP utilised the overlap-and-save method\cite{Oppenheim:book1975} for the processing of the incoming data streams, with overlap length and FFT size of $1024$ and $2048$ samples, respectively. These values were optimised to ensure performance and avoid penalties across all MIMO dimensions. The receiver's DSP blocks were implemented as a differentiable computation graph in TensorFlow. During the training
phase, the outputs of the MIMO NLE were linked to a single mean-squared-error function for computing the gradients of the model's trainable parameters, 
$L_{\mathrm{MSE}}=\frac{1}{N_{\mathrm{ch}} K} \sum_{n=1}^{N_{\mathrm{ch}}}\sum_{c=1}^{K}\abs{s_{\mathrm{out},n}^{(c)}-\hat{s}_{\mathrm{out},n}^{(c)}}^2,
$
where $\hat{s}_{\mathrm{out},n}^{(c)}$ and $s_{\mathrm{out},n}^{(c)}$ are the reference and recovered symbols, respectively, and $K$ is the total number of symbols within a
batch.
During the testing phase, the recovered symbols from each channel were used to compute the bit error rate, which was then mapped to an effective signal-to-noise ratio (SNR).
For a given launch power, datasets included $2^{19}$ symbols for training and $2^{18}$ symbols each for validation and testing.
T,he model was trained using the Adam optimiser, with a learning rate of $0.001$ and a batch size of $40$. MIMO models of varying sizes were trained separately for each launch power. Training was done over $750$ epochs, after which no further improvements were observed.

\begin{figure*}[t]
    \centering
    \includegraphics[height=5.6cm]{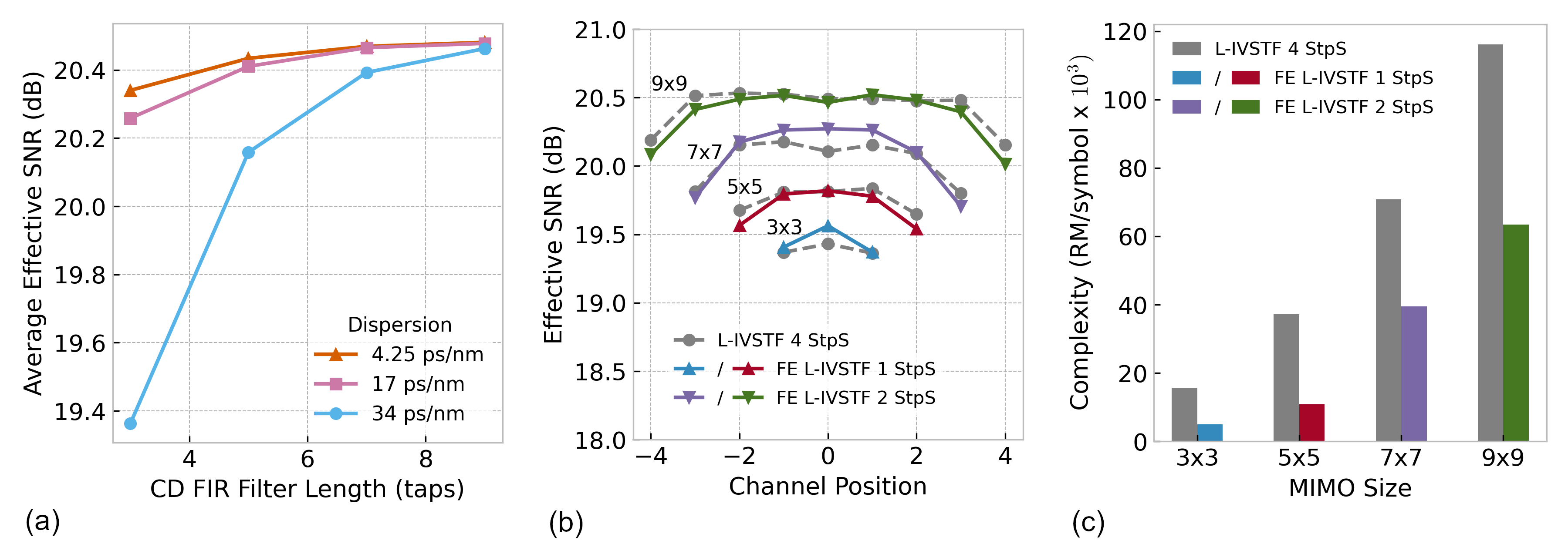}
    \caption{(a) Average SNR performance versus length of the CD FIR filters for the $9\times 9$ MIMO FE L-IVSTF scheme operating at $N_{\mathrm{StpS}}=2$. (b) Per-channel performance of different-sized MIMO FE L-IVSTF schemes using $3$-tap CD FIR filters, operating at $N_{\mathrm{StpS}}=1$ ($3\times 3$, $5\times 5$) and $N_{\mathrm{StpS}}=2$ ($7\times 7$, $9\times 9$), and of the corresponding L-IVSTF schemes operating at $N_{\mathrm{StpS}}=4$. (c) Computational cost of the FE L-IVSTF and L-IVSTF models in (b).}
    \label{fig:figure2}
\end{figure*}

Our model's trainable parameters included the coefficients $\alpha_c^{(k)}$, $\beta_{c,r}^{(k)}$ of the SPM and XPM FIR filters, respectively, and the coefficients of the $h_{i,n}^{(k)}$ and $h_{o,n}^{(k)}$ CD FIR filters. The SPM and XPM filters were initialised with zero-valued taps, and the CD filters were initialised following the method in \cite{Sheikh_lsco}. The hyper-parameters of the model, including the lengths of all FIR filters and the amount of dispersion to be compensated by the CD FIR filters,  were optimised to maximise performance and reduce complexity. Figure \ref{fig:figure2}(a) shows the average effective SNR of the channels as a function of the length $S_{\mathrm{CD}}$ of the CD FIR filters for varying amounts of dispersion for the $9\times 9$ MIMO model operated at two steps per span. The optimum launch channel power was used, with the lengths of all SPM and XPM filters across the structure set at $S_{\mathrm{SPM}}=7$ and $S_{\mathrm{XPM}}=31$ taps, respectively. We can see that a FIR-compensated dispersion of $4.25\,\mathrm{ps/nm}$ is sufficient to achieve the desired adaptability for CD compensation,
hence sufficient inter-channel equalisation. 
For this low dispersion value, the optimum FIR filter's length beyond which performance saturates is $7$ taps. Choosing shorter filters results in performance penalties of up to $0.2\,\mathrm{dB}$.   

Figure \ref{fig:figure2}(b) shows the SNR performance for each processed channel in different-sized MIMO FE L-IVSTF implementations using $3$-tap CD FIR filters. For comparison, the performance of L-IVSTF models without FE filtering is shown. The results demonstrate that the $3\times3$ and $5\times5$ FE L-IVSTF configurations match the performance of their L-IVSTF counterparts at just one step per span, while the $7\times7$ and $9\times9$ configurations require two steps per span to produce a similar outcome. It is noteworthy that selecting slightly larger CD FIR filter's lengths ($5$ or $7$ taps) can bring some performance gain without sacrificing the computational complexity. 

To contextualise these findings, we conducted a complexity analysis focusing on signal-path
real multiplications (RMs) of constant resolution per transmitted symbol. Starting with 
FD operations, the FE L-IVSTF model includes $N_s+1$ pairs of forward and inverse FFT operations, where $N_s= N_{\mathrm{StpS}} N_{\mathrm{sp}}$ represents the total number of steps. Each FFT pair, with a radix-2 implementation, incurs a cost of $C_{\mathrm{FFT}} = 4N_{\mathrm{FFT}}\log_{2}(N_{\mathrm{FFT}})$ RMs for a sample block of length $N_{\mathrm{FFT}}$. Additionally, the element-wise complex multiplications between the transformed signal and the FD CD filters require $4N_{\mathrm{FFT}}$ RMs per linear step. Therefore, the total cost of the FD operations is $C_{\mathrm{MIMO}, \mathrm{FD}} = [q(1+N_{s})C_{\mathrm{FFT}}+(4qN_{\mathrm{FFT}})(2N_{s}+1)]/(N_{\mathrm{FFT}}-M+1)
$ RMs per symbol (RM/sym), where $q$ is the digital sampling rate, and $M$ is the overlap length. For TD operations, the SPM and XPM filtering of signal powers 
requires $0.5q(S_{\mathrm{SPM}}+1)$ and $qS_{\mathrm{XPM}}$ RM/sym, respectively, where $(N_{\mathrm{ch}}-1)$ XPM filtering operations are needed for each nonlinear activation.
The nonlinear activation functions add another $4q$ RM/sym due to squared signal modules and multiplications by complex constants. The CD FIR filters add $8q S_{\mathrm{CD}}$ RM/sym for each nonlinear branch of the FE L-IVSTF model, leading to the total cost for TD operations of:
$C_{\mathrm{MIMO}, \mathrm{TD}} = qN_{ch} N_{s}[0.5(S_{\mathrm{SPM}}+1)+(N_{\mathrm{ch}}-1)S_{\mathrm{XPM}} + 8 S_{\mathrm{CD}}+4]$ RM/sym. The results are summarised in Fig.\ \ref{fig:figure2}(c), which illustrates the total computational complexity ($C_{\mathrm{MIMO,FD}}+C_{\mathrm{MIMO,TD}}$) imposed on the L-IVSTF and FE L-IVSTF models to achieve the same performance (as depicted in Fig.\ \ref{fig:figure2}(b)). Although the per-step complexity of the FE L-IVSTF model is marginally higher than that of its L-IVSTF counterpart, its ability to operate with fewer steps results in a significant overall reduction in complexity. The most substantial reduction is observed in the $3 \times 3$ MIMO configuration, with a total cost of approximately $4935.24$ RM/sym, representing $31.51\%$ of the L-IVSTF model's complexity. Conversely, the $9 \times 9$ configuration requires $63374.59$ RM/sym, representing $54.55\%$ of the multiplications needed by the corresponding L-IVSTF model. 


\section{Conclusions}
We have proposed a novel learned MIMO Volterra-based scheme, demonstrating 
its effective operation with a minimal number of computational steps without sacrificing performance. This structure, featuring adaptive nonlinear stages enhanced by filtering both the power and optical signal waveforms, achieves significant complexity reduction with a $46\%$ decrease for a $9\times9$ implementation compared to its counterpart with only power waveform filtering.

\section{Acknowledgements}
This work was partly supported by the UK EPSRC grants TRANSNET (EP/R035342/1), CREATE (EP/X019241/1), and EEMC (EP/S016171/1).
\defbibnote{myprenote}{

}
\printbibliography[prenote=myprenote]

\vspace{-4mm}

\end{document}